\documentclass[prl,twocolumn,superscriptaddress,showpacs,amsmath]{revtex4}
\date{}
\usepackage{graphicx}

\begin{document}
\title{First-principles energetics of water: a many-body analysis}

\author{M. J. Gillan}
\affiliation{London Centre for Nanotechnology, UCL, London WC1H 0AH, UK}
\affiliation{Dept. of Physics and Astronomy, UCL, London WC1E 6BT, UK}
\affiliation{Thomas Young Centre, UCL, London WC1H 0AH, UK}

\author{D. Alf\`{e}}
\affiliation{Dept. of Physics and Astronomy, UCL, London WC1E 6BT, UK}
\affiliation{Thomas Young Centre, UCL, London WC1H 0AH, UK}
\affiliation{Dept. of Earth Sciences, UCL, London WC1E 6BT, UK}

\author{A. P. Bart\'{o}k}
\affiliation{Dept. of Engineering, University of Cambridge,  Cambridge, UK}

\author{G. Cs\'{a}nyi}
\affiliation{Dept. of Engineering, University of Cambridge, Cambridge, UK}


\begin{abstract}
Standard forms of density-functional theory (DFT) have good predictive power
for many materials, but are not yet fully satisfactory for solid, liquid and cluster forms of 
water. We use a many-body separation of the total energy into its 1-body, 2-body (2B) and beyond-2-body 
(B2B) components to analyze the deficiencies of two popular DFT approximations. We show
how machine-learning methods make this analysis possible
for ice structures as well as for water clusters. We find that the crucial
energy balance between compact and extended geometries can be distorted by
2B and B2B errors, and that both types of first-principles error are important. 

\pacs{31.15.E-, 61.50.Lt, 71.15.Mb}

\end{abstract}

\maketitle

\clearpage

The pioneering work of Parrinello, Car and others in the 
1990s~\cite{laasonen93} initiated a major effort to
understand the properties of water from first principles using density-functional theory (DFT). This
effort is important not just for pure water, but for general aqueous systems,
including solutions~\cite{laasonen96} and water on surfaces~\cite{lindan98}. 
However, standard DFT methods often give less than satisfactory
predictions for water in its liquid~\cite{grossman04}, 
crystalline~\cite{hamann97,santra11} and cluster 
forms~\cite{anderson06,gillan12,santra08}, for reasons that are
controversial. Strenuous efforts have been made
to overcome the problems by adding correction terms, with a recent emphasis on
correcting dispersion (see e.g.
Refs~\cite{schmidt09,wang11,ma12,kelkkanen09,mogelhoj11,santra11}). 
Here we show how a combination of
quantum chemistry, machine learning and quantum Monte Carlo can be used to analyze the energetics of
water in a variety of aggregation states. We find that DFT approximations often distort
the energy balance between extended and compact structures, and that errors can arise from
more than one part of the first-principles energy. Technical details of our calculations are given in the
Supplemental Material~\cite{SI}.

Our starting point is that a model for the energetics of water is not fully satisfactory unless
it gives good predictions for a range of aggregation states, and particularly: water clusters  (including
the dimer) in a variety of geometries, ice structures, and the liquid. By this criterion,
standard DFT approximations need improvement, since their many errors
include: inaccurate predictions of energies for some dimer 
geometries~\cite{anderson06,gillan12}; wrong stability ordering
of isomers of some clusters, notably the hexamer~\cite{santra08,dahlke08}; incorrect relative
energies of different ice structures~\cite{santra11}; errors of up to $30$~\% in the predicted density of
the liquid~\cite{mcgrath05,schmidt09,wang11,ma12}; and substantial errors in the 
structure and diffusivity of the liquid~\cite{grossman04,schmidt09,wang11,mogelhoj11,ma12}. 

Our analysis of water energetics is based on the
many-body decomposition, in which the total energy $E_{\rm tot} ( 1, 2, \ldots N )$ of a
system of $N$ monomers is expressed as~\cite{xantheas94}:
\begin{equation}
\begin{split}
E_{\rm tot} ( 1, 2, \ldots N ) & =  \sum_i E^{(1)} ( i ) + \\
& \sum_{i < j} E^{(2)} ( i, j ) + E^{(>2)} ( 1, 2, \ldots N ) \; .
\end{split}
\end{equation}
Here, $i$ is the collection of variables describing monomer $i$ (position, orientation, internal
distortion from equilibrium geometry), $E^{(1)} ( i )$ the 1-body (1B) energy of monomer $i$,
$E^{(2)} ( i, j )$ the 2-body (2B) interaction energy of monomers $i$ and $j$, and the 
beyond-2-body (B2B) energy $E^{(>2)}$ is everything not included in 1B and 2B energy. 
In this scheme, dispersion (non-local correlation) is 
mainly a 2B energy~\cite{stone96,klimes12},
though it also contributes to the B2B energy~\cite{wang10}. Induction is usually
regarded as the largest contributor to B2B, though many-body
exchange-repulsion may also be significant~\cite{wang10}.

\begin{figure}[htb]
\centerline{
\includegraphics[width=0.7\linewidth]{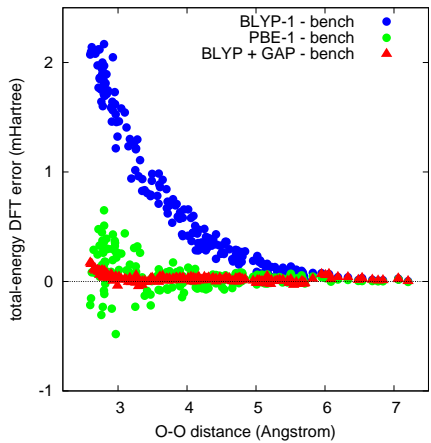}
}
\caption{Errors of 2-body energy of H$_2$O dimer with BLYP and PBE functionals
relative to CCSD(T) benchmarks
as function of O-O distance. Calculations are for sample of 198 dimer
configurations drawn from a classical m.d. simulation of bulk water at
ambient conditions. Also shown are residual errors of the approximation
obtained by adding GAP 2-body corrections to BLYP. Units: m$E_{\rm h}$.}
\label{fig:dimer_dft_errors}
\end{figure}

We start with the water dimer, whose energy in any geometry can be accurately computed
(within $0.1$~m$E_{\rm h} \simeq$ 
$2$~meV~$\simeq$ $0.05$kcal/mol~\cite{tschumper02}) using 
the correlated quantum chemistry 
technique of CCSD(T) (coupled cluster single and double excitations 
with a perturbative treatment of triples)~\cite{helgaker00}. We take the difference 
$\delta E^{(2)} ( \mathrm{DFT} ) \equiv E^{(2)} ( \mathrm{DFT} ) - E^{(2)} ( \mathrm{CCSD(T)} )$
as the error of any DFT approximation for the 2B energy of any dimer geometry.
Following Refs.~\cite{santra09,gillan12}, we study the errors
$\delta E^{(2)} ( \mathrm{DFT} )$ for thermal samples of dimer geometries randomly drawn
from an m.d. simulation of the liquid, the simulation used here being
done with the classical \textsc{amoeba} force field~\cite{ren03}, as in Ref.~\cite{gillan12}. 
Fig.~\ref{fig:dimer_dft_errors} shows the 2B errors of the widely
used PBE~\cite{perdew96} and BLYP~\cite{becke88} approximations \emph{versus} 
O-O distance for a sample of 198 dimer
configurations~\cite{SI}, demonstrating  (see also Refs.~\cite{gillan12,santra07}) 
that BLYP systematically underbinds for all distances and
molecular orientations, while PBE performs better (but see Ref.~\cite{ireta2004}). We 
have shown recently~\cite{bartok13} that machine-learning
methods based on the ideas of Gaussian approximation potentials (GAP)~\cite{bartok10} can be used to
represent accurately (within $0.1$~m$E_{\rm h} \simeq 2$~meV 
or better) the 1B and 2B errors of chosen DFT
approximations in water systems. We illustrate this by including
in Fig.~\ref{fig:dimer_dft_errors} the tiny residual 2B errors of GAP-corrected BLYP~\cite{SI}.
This gives a way of correcting almost perfectly for the 2B errors of any
DFT approximation.

\begin{figure}[htb]
\begin{tabular}{cc}
\includegraphics[width=0.5\linewidth]{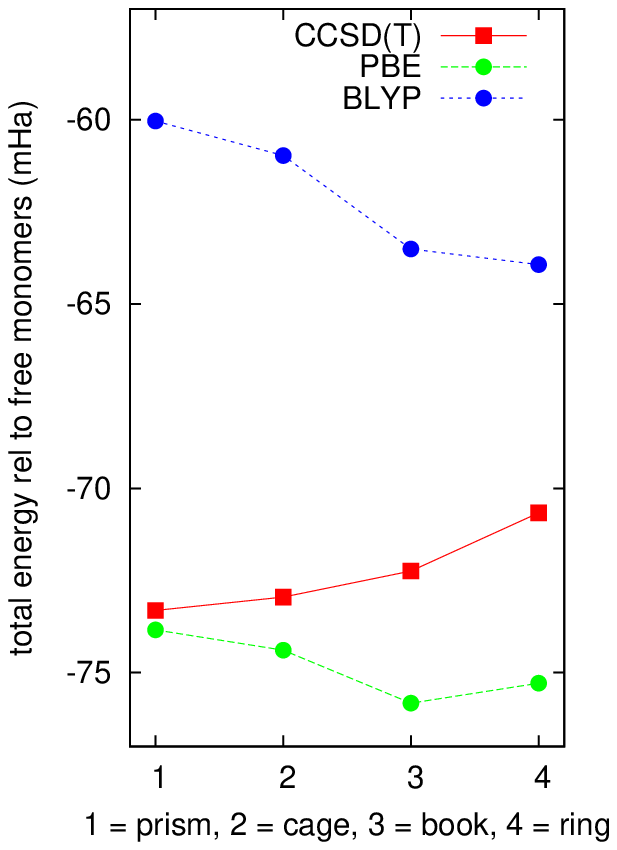} &
\includegraphics[width=0.5\linewidth]{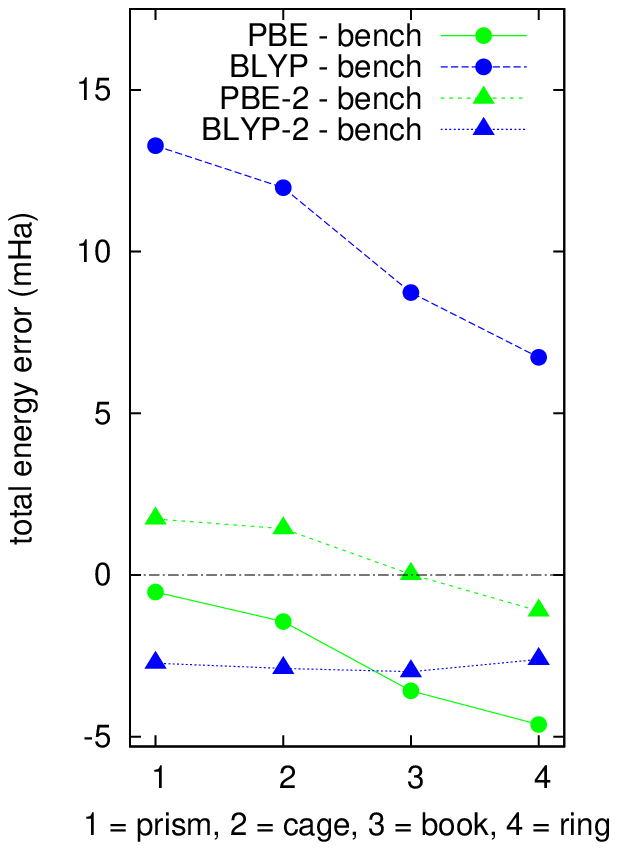}
\end{tabular}
\caption{Left panel: total energies of isomers of the H$_2$O hexamer relative to free monomers
from PBE, BLYP and benchmark CCSD(T) calculations; right panel: errors of total energy
of PBE and BLYP and 1- and 2-body corrected PBE-2 and BLYP-2. Units: m$E_{\rm h}$}
\label{fig:hex_isomers}
\end{figure}

The ability to correct almost exactly for 1B and 2B errors in the total energy of any water system is
invaluable, because it lets us decompose DFT error into its 1B, 2B and B2B components.
We show how this works for the isomers of the water hexamer. The energies of the prism, cage, book
and ring isomers of the hexamer (see e.g. Ref.~\cite{santra08} for pictures)  have been intensively 
studied~\cite{bates09,dahlke08,santra08,kim94} for an important reason. For smaller
clusters, the most stable isomers have ring-like geometries, but from the hexamer onwards compact
geometries are more stable. Highly converged CCSD(T) calculations~\cite{bates09} show that the compact prism
and cage are more stable than the extended book and ring isomers. However, 
standard DFT approximations wrongly make the extended geometries more stable~\cite{santra08}: our
own calculations of the total energies (Fig.~\ref{fig:hex_isomers}) illustrate this for PBE 
and BLYP~\cite{SI}. If we now correct
for 1B and 2B errors by adding the GAP representation of the differences
$\mathrm{DFT} - \mathrm{CCSD(T)}$, then the errors of the resulting approximations (we denote
them by PBE-2 and BLYP-2) are by definition B2B errors. As shown in Fig.~\ref{fig:hex_isomers}, the errors
of BLYP-2 are negative but almost constant, so that the relative energies are now excellent.
The errors of PBE-2 are smaller than those of PBE itself, but are still not negligible.
This means that the erroneous stability ordering with BLYP is mainly due to 2B effects, but with PBE
both 2B and B2B effects are  important, as pointed out in 
earlier work~\cite{wang10,gillan12}. 

\begin{figure}[htb]
\begin{tabular}{cc}
\includegraphics[width=0.5\linewidth]{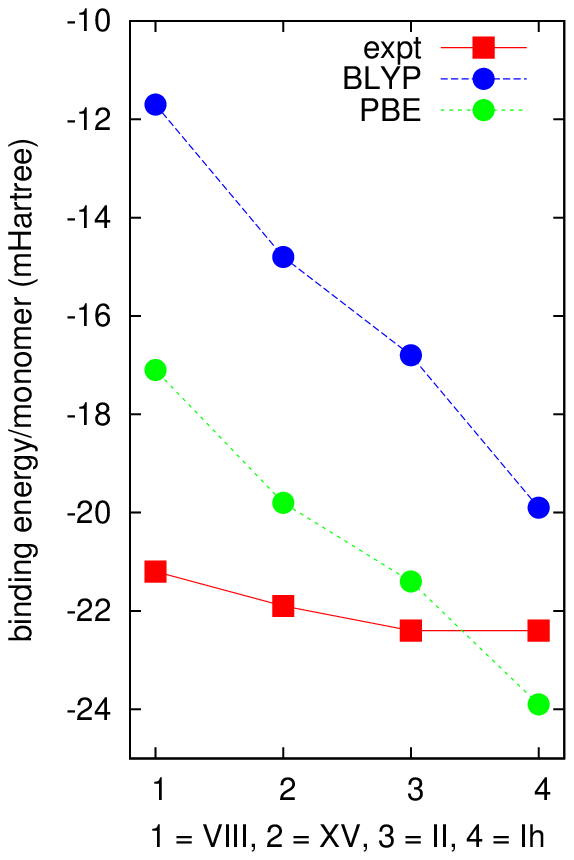} &
\includegraphics[width=0.5\linewidth]{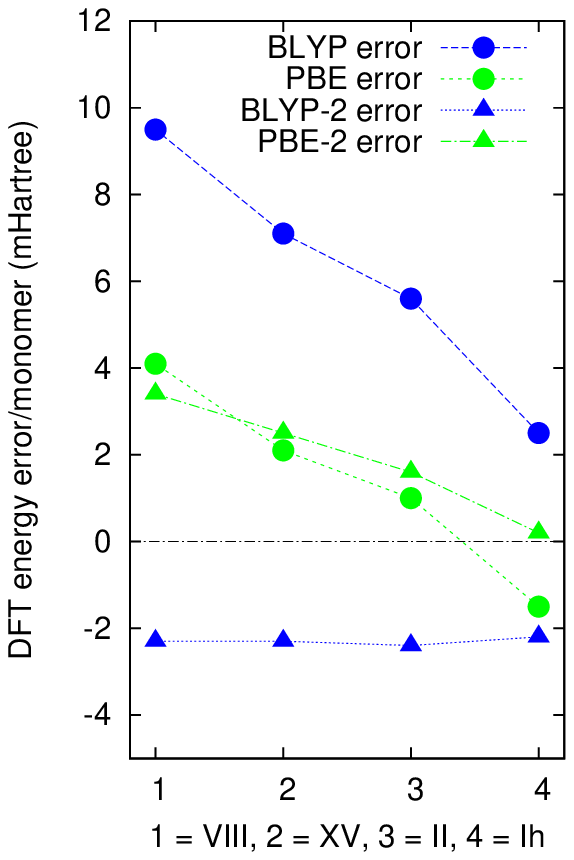}
\end{tabular}
\caption{Left panel: energies of ice structures relative to free monomers
from PBE, BLYP and experiment; right panel: errors of PBE and BLYP
and 1- and 2-body corrected PBE-2 and BLYP-2 energies. Energy units: m$E_{\rm h}$.}
\label{fig:ice_structures}
\end{figure}

We turn now to the energetics of ice structures, which gives striking evidence of the difficulties
of standard first-principles methods~\cite{santra11}. Essentially the same analysis that we used for the hexamers
helps determine the nature of DFT errors for the cohesive energies of ice structures. Ice has
a complicated phase diagram, with no less than 15 known crystal structures~\cite{petrenko99}, but we study only
the proton-ordered structures XI, II, XV and VIII forming the sequence of increasingly
compact  structures found at low
temperatures as pressure increases from $0$ to $\sim 20$~kbar. The errors of DFT for these and other
structures have recently been studied in detail~\cite{santra11}, and it was shown that the predicted  energies
increase much too fast from extended to compact structures. We 
illustrate this in Fig.~\ref{fig:ice_structures},
where our calculated cohesive energies with PBE and BLYP~\cite{SI} are compared with experimental 
values~\cite{whalley84} (zero-point energies removed, see also Ref.~\cite{santra11}). 

The GAP 1B and 2B corrections to PBE and BLYP are readily computed in periodic
boundary conditions~\cite{bartok13}, and Fig.~\ref{fig:ice_structures} shows 
the errors of the uncorrected and corrected approximations
for the XI, II, XV and VIII structures. The picture resembles what we saw for the hexamers, with BLYP being 
increasingly underbinding as we go from extended to compact structures, but its corrected version
BLYP-2 having almost constant negative errors, so that its relative energies are very good.
By contrast, the corrected version PBE-2, while better than PBE itself, still gives substantial errors.
This implies that for BLYP the problem with relative energies stems mainly from 
systematically underbinding 2B interaction,
but that B2B errors are also important for PBE.

The energy changes with increasing compactness along the series XI, II, XV, VIII can be understood
in more detail. In all these structures, each H$_2$O monomer is hydrogen-bonded to four
first neighbors at O-O distances of $\sim 2.7$~\AA~\cite{petrenko99}. In XI (the proton-ordered form of the
Ih structure of common ice) the monomers form a tetrahedral network, the second-neighbors being
at the large distance of $4.5$~\AA. From XI to II, XV and VIII, the hydrogen-bonded
1st-neighbor distances change only slightly, but the 2nd-neighbor distances contract
dramatically, until in VIII each monomer has eight neighbors at almost equal distances of $\sim 2.8$~\AA,
four of which are unbonded to the central monomer~\cite{petrenko99}. The close approach of monomers that are not
H-bonded to each other in the compact structures appears to be implicated in the large DFT errors,
as has been pointed out before (e.g. Ref.~\cite{wang11}).

\begin{figure}[htb]
\begin{tabular}{cc}
\includegraphics[width=0.5\linewidth]{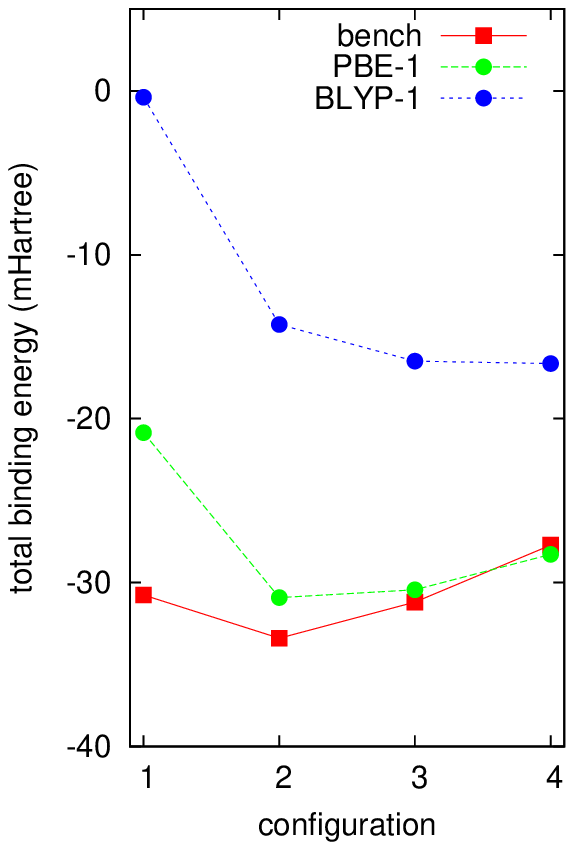} &
\includegraphics[width=0.5\linewidth]{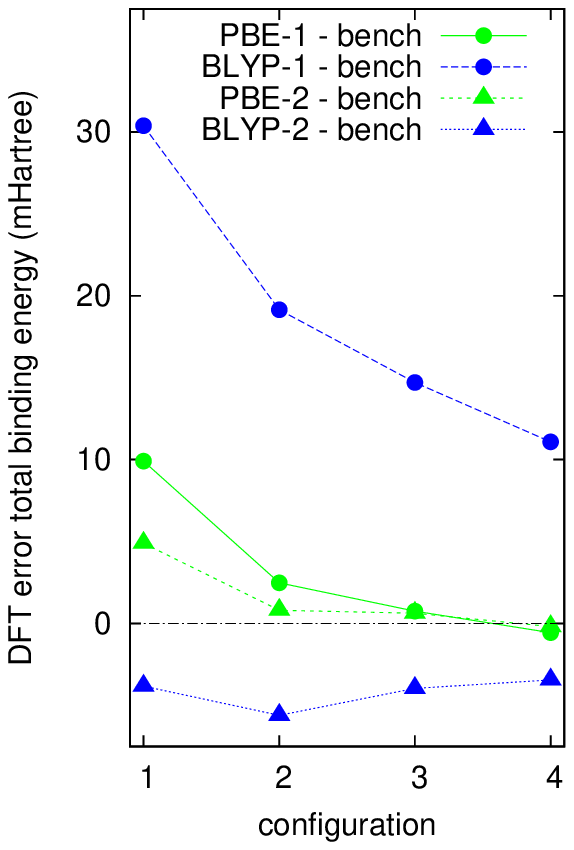} 
\end{tabular}
\caption{Left panel: total binding energies of the four nonamers derived from the ice VIII
structure computed with DMC, PBE-1 and BLYP-1; non-bonded neighbors move radially
outward as we go from left to right. Right panel: errors of the total energy relative to DMC
with PBE-1, BLYP-1, PBE-2 and BLYP-2 DFT approximations. Energy units: m$E_{\rm h}$.}
\label{fig:errors_iceVIII_nonamer}
\end{figure}

To further probe the DFT errors in the
denser ice structures, we cut from ice VIII the nonamer composed of an H$_2$O molecule
and its nearest neighbors, and we study the energy changes when the H-bonded
neighbors are held fixed but the unbonded neighbors are moved radially. We calculated benchmark
energies for the resulting configurations using diffusion Monte 
Carlo~\cite{SI,foulkes01}, which is extremely
accurate both for water clusters and for ice structures~\cite{santra08,santra11,gillan12}, 
including ice VIII. Comparison with PBE and BLYP
energies~\cite{SI} (Fig.~\ref{fig:errors_iceVIII_nonamer}) shows that both 
approximations give excessive energy increases on going from
extended to compact configurations. Comparing the GAP-corrected approximations
PBE-2 and BLYP-2 with uncorrected PBE and 
BLYP (Fig.~\ref{fig:errors_iceVIII_nonamer}), we see again what we learnt
from the hexamers and the ice structures. Correction for 1B and 2B errors takes BLYP
from severely underbinding to somewhat overbinding, but with almost constant B2B errors; 
corrected PBE, while better than uncorrected PBE, still suffers from similar (though smaller)
errors, so that B2B effects are important.

Several key points emerge from our analysis. First, the pervasive importance of the energy balance between
extended and compact structures is highlighted by the fact that we see the same pattern of
DFT errors in the hexamers, the ice structures and the nonamer configurations. Second, we have seen that
both 2B and B2B errors can distort this balance. With the BLYP functional, 2B errors are the main
culprit in tipping the balance towards extended structures, but with PBE the contribution of B2B errors is significant. Third, it is clear that different kinds of B2B errors are important. With PBE,
the B2B errors appear to be associated with the close approach of monomers that are not H-bonded to
each other (ice structures, nonamer) or that involve highly distorted H-bonds (hexamers). With BLYP,
by contrast, B2B overbinding occurs even for extended structures, and may be due to exaggerated
cooperativity of H-bonding. Since dispersion is expected to be mainly a 2B interaction in water, our results
clearly indicate that sources of error in addition to the poor treatment of dispersion should be considered.

Is our analysis relevant to the first-principles understanding of the density, structure
and diffusivity of liquid water? It seems likely that the systematic 2B underbinding of
BLYP accounts both for its under-prediction of the equilibrium density and for its prediction of an
over-structured and under-diffusive liquid; however, its B2B overbinding might also be expected
to contribute to over-structuring and under-diffusivity. On the other hand, the under-prediction of the
density by PBE may arise from the excessive 2B and B2B repulsion involving H$_2$O monomers
that are not H-bonded to each other, an effect that could also account for over-structuring and
under-diffusivity. These suggestions may be directly testable, since diffusion Monte Carlo
should be able to provide accurate total-energy benchmarks for the liquid in periodic boundary conditions; if so,
the many-body analysis given here would be feasible for thermal samples of the liquid.

Our analysis of first-principles errors for a variety of water systems 
into 1-, 2- and beyond-2-body components gives helpful insights into their fundamental energetics,
but a detailed energy decomposition analysis (see e.g. Ref.~\cite{wang10}) is also needed,
including an assessment of the role of non-local electron correlation in
the energetics of the dimer and the other water systems studied here.


\acknowledgments{APB was supported by a Junior Research Fellowship at Magdalene College, Cambridge.
GC was supported by the Office of Naval Research under grant N000141010826, and by
EU-FP7-NPM grant 229205 ADGLASS.
We used resources of the Oak Ridge Leadership Computing Facility in the National Center
for Computational Sciences at ORNL, 
which is supported by the Office of Science of the DOE under Contract No.
DE-AC05-00OR22725. Access to HECToR, the UK's national
high-performance computing service is acknowledged. We thank A. Michaelides for comments on the manuscript.}



\begin{thebibliography}{99}
\bibitem{laasonen93}
K. Laasonen, M. Sprik, M. Parrinello, and R. Car, J. Chem. Phys. \textbf{99}, 9080 (1993); C. Lee,
D. Vanderbilt, K. Laasonen, R. Car, and M. Parrinello, Phys. Rev. B, \textbf{47}, 4863 (1993);
M. E. Tuckerman, K. Laasonen, M. Sprik, and M. Parrinello, J. Phys. Condens. Matter, \textbf{6},
A93 (1994); M. Sprik, J. Hutter, and M. Parrinello, J. Chem. Phys., \textbf{105}, 1142 (1996).

\bibitem{laasonen96}
See e.g. K. Laasonen and M. L. Klein, Mol. Phys. \textbf{88}, 135 (1996); E. J. Meijer and M. Sprik,
J. Amer. Chem. Soc. \textbf{120}, 6345 (1998); J. Blumberger, L. Bernasconi, I. Tavernelli,
R. Vuilleumier, and M. Sprik, J. Amer. Chem. Soc. \textbf{126}, 3928 (2004).

\bibitem{lindan98}
See e.g. P. J. D. Lindan, N.M. Harrison, and M. J. Gillan, Phys. Rev. Lett. \textbf{80}, 762 (1998); 
L. Liu, M. Krack, and A. Michaelides, J. Amer. Chem. Soc. \textbf{130}, 8572 (2008); J. Carrasco,
A. Hodgson, and A. M. Michaelides, Nature Mater. \textbf{11}, 667 (2012).

\bibitem{grossman04}
J. C. Grossman, E. Schewegler, E. W. Draeger, F. Gygi, and G. Galli, J. Chem. Phys. \textbf{120} 300 (2004);
M. Allesch, E. Schwegler, F. Gygi, and G. Galli, J. Chem. Phys. \textbf{120}, 5192 (2004);
P. H.-L. Sit and N. Marzari, J. Chem. Phys. \textbf{122}, 204510 (2005).

\bibitem{hamann97}
D. R. Hamann, Phys. Rev. B, \textbf{55}, R10157 (1997).

\bibitem{santra11}
B. Santra, J. Klime\v{s}, D. Alf\`{e}, A. Tkatchenko, B. Slater, A. Michaelides, R. Car, and M. Scheffler,
Phys. Rev. Lett., \textbf{107}, 185701 (2011); J. Klime\v{s}, D. R. Bowler, and A. Michaelides, J. Phys. Condens. 
Matter, \textbf{22}, 022201 (2010).

\bibitem{anderson06}
J. A. Anderson and G. S. Tschumper, J. Phys. Chem. A, \textbf{110}, 7268 (2006).

\bibitem{gillan12} 
M. J. Gillan, F. R. Manby, M. D. Towler, and D. Alf\`{e}, J. Chem. Phys., \textbf{136}, 244105 (2012).

\bibitem{santra08}
B. Santra, A. Michaelides, M. Fuchs, A. Tkatchenko, C. Filippi, and M. Scheffler, J. Chem. Phys., \textbf{129},
194111 (2008).

\bibitem{schmidt09}
J. Schmidt, J. VandeVondele, I.-F. W. Kuo,
D. Sebastiani, J. I. Siepmann, J. Hutter, C. J. Mundy, J. Phys. Chem. B, \textbf{113}, 11959 (2009)

\bibitem{kelkkanen09}
A. K. Kelkkanen, B. I. Lundqvist, and J. K. Norskov, J. Chem. Phys., \textbf{131}, 046102 (2009).

\bibitem{wang11}
J. Wang, G. Rom\'{a}n-P\'{e}rez, J. M. Soler, E. Artacho, and M.-V. Fern\'{a}ndez-Serra,
J. Chem. Phys., \textbf{134}, 024516 (2011).

\bibitem{mogelhoj11}
A. Mogelhoj et al., J. Phys. Chem. B, \textbf{115}, 14149 (2011).

\bibitem{ma12}
Z. Ma, Y. Zhang, and M. E. Tuckerman, J. Chem. Phys., \textbf{137}, 044506 (2012).

\bibitem{SI}
See Supplemental Material at (URL here) for computional details.

\bibitem{dahlke08}
E. E. Dahlke, R. M. Olson, H. R. Leverentz, and D. G. Truhlar, J. Phys. Chem. A,
\textbf{112}, 3976 (2008).

\bibitem{kim94}
K. Kim, K. D. Jordan, and T. S. Zwier, J. Amer. Chem. Soc. \textbf{116}, 11568 (1994).

\bibitem{mcgrath05}
M. J. McGrath, J. I. Siepmann, I.-F. W. Kuo, C. J. Mundy, J. VandeVondele, J. Hutter, F. Mohamed, and
M. Krack, Chem. Phys. Chem. \textbf{6}, 1894 (2005).

\bibitem{xantheas94}
S. S. Xantheas, J. Chem. Phys., \textbf{110}, 7523 (1994); J. M. Pedulla, F. Vila, and K. D. Jordan,
J. Chem. Phys., \textbf{105}, 11091 (1996).

\bibitem{stone96}
A. J. Stone, \emph{Theory of Intermolecular Forces} (Oxford University Press, Oxford, 1996).

\bibitem{klimes12}
J. Klime\v{s} and A. Michaelides, J. Chem. Phys., \textbf{137}, 120901 (2012).

\bibitem{wang10}
F.-F. Wang, G. Jenness, W. A. Al-Saidi, and K. D. Jordan, J. Chem. Phys. \textbf{132}, 134303 (2010).

\bibitem{tschumper02}
G. S. Tschumper, M. L. Leininger, B. C. Hoffman, E. F. Valeev, H. F. Schaeffer III, and M. Quack,
J. Chem. Phys., \textbf{116}, 690 (2002).

\bibitem{helgaker00}
T. Helgaker, P. Jorgensen, and J. Olsen, \emph{Molecular Electronic-Structure Theory}
(Wiley, New York, 2000).

\bibitem{santra09}
B. Santra, A. Michaelides, and M. Scheffler, J. Chem. Phys., \textbf{131}, 124509 (2009).

\bibitem{ren03}
P. Ren and J. W. Ponder, J. Phys. Chem. B, \textbf{107}, 5933 (2003).

\bibitem{perdew96}
J. Perdew, K. Burke, and M. Ernzerhof, Phys. Rev. Lett., \textbf{77}, 3865 (1996).

\bibitem{becke88}
A. D. Becke, Phys. Rev. A, \textbf{38}, 3098 (1988); C. Lee, W. Yang, and R. Parr, Phys. Rev. B,
\textbf{37}, 785 (1988).

\bibitem{santra07}
B. Santra, A. Michaelides, and M. Scheffler, J. Chem. Phys., \textbf{127}, 184104 (2007).

\bibitem{ireta2004}
J. Ireta, J. Neugebauer, and M. Scheffler, J. Phys. Chem. A, \textbf{108}, 5692 (2004)

\bibitem{bartok13}
A. P. Bart\'{o}k, M. J. Gillan, F. R. Manby, and G. Csanyi, Proc. Natl. Acad. Sci. USA, submitted. (Arxiv cond-mat 1302.5680.)

\bibitem{bartok10}
A. P. Bart\'{o}k, M. C. Payne, R. Kondor, and G. Cs\'{a}nyi, Phys. Rev. Lett., \textbf{104}, 136403 (2010).

\bibitem{bates09}
D. M. Bates and G. S. Tschumper, J. Phys. Chem. A, \textbf{113}, 3555 (2009).

\bibitem{petrenko99}
V. F. Petrenko and R. W. Whitworth, \emph{Physics of Ice}, Oxford University Press (1999);
C. G. Salzmann, P. G. Radaelli, E. Mayer, and J. L. Finney, Phys. Rev. Lett. \textbf{103}, 105701 (2009).

\bibitem{whalley84}
E. Whalley, J. Chem. Phys., \textbf{81}, 4087 (1984).

\bibitem{foulkes01}
See e.g. W. M. C. Foulkes, L. Mita\v{s}, R. J. Needs, and G. Rajagopal, Rev. Mod. Phys., \textbf{73}, 33 (2001);
R. J. Needs, M. D. Towler, N. D. Drummond, and P. L\'{o}pez-R\'{\i}os, J. Phys. Condens. Matter, \textbf{22},
023201 (2010).

\end{thebibliography}
\end{document}